\pdfoutput=1
\documentclass[journal=jacsat,manuscript=communication]{achemso}

\usepackage[version=3]{mhchem} 
\usepackage{epstopdf} 
\usepackage[dvipsnames]{xcolor}
\usepackage{MnSymbol}

\author{Millan Michelin-Jamois}
\affiliation[INSA-Lyon]
{MATEIS, INSA-Lyon, CNRS UMR 5510, 7 avenue Jean Capelle, 69621 Villeurbanne Cedex, France}
\author{Cyril Picard}
\author{Elisabeth Charlaix}
\affiliation[Universite Joseph Fourier]
{LIPHY, Universite Joseph Fourier, CNRS UMR 5588, 140 Avenue de la Physique, 68402 Saint Martin d'Heres, France}
\email{Elisabeth.Charlaix@ujf-grenoble.fr}
\phone{+33 (0)476 514 963}
\fax{+33 (0)476 635 495}
\author{Gerard Vigier}
\affiliation[INSA-Lyon]
{MATEIS, INSA-Lyon, 7 avenue Jean Capelle, 69621 Villeurbanne Cedex, France}

\title[An \textsf{achemso} demo]
  {Intrusion of liquid electrolytes inside hydrophobic MOF's : contribution of the osmotic pressure }

\abbreviations{LHS, ZIF-8}
\keywords{Osmotic pressure, lyophobic heterogeneous systems (LHS), nucleation, water intrusion, electrolytes}

\begin{document}

%
%

\begin{abstract}
{High pressure water intrusion in hydrophobic nanoporous media have been studied in relation with energy storage. It has already been showed that addition of electrolytes in water increases intrusion pressure of liquid leading to an enhancement of storing capacities. We demonstrate here that for a number of salt, a very simple van 't Hoff law can explain intrusion and extrusion excess pressures compared with the pure water case. Our results suggest that only pure water can penetrate the pores, the ions being quartered in the bulk liquid around nanoporous medium. This selectivity explains very high pressures for very concentrated ions reported elsewhere \cite{Tzanis2014}. Finally, a partial intrusion of $NaI$ and $LiI$ is observed. This effect could be explained on the basis of particular behaviour or iodide ions over hydrophobic surfaces.}

\end{abstract}

\section{Introduction}

Heterogeneous Lyophobic systems (HLS), based on forced intrusion and spontaneous extrusion cycles of a nonwetting fluid in a nano-porous media, are particularly appealing for either storage or dissipation of mechanical energy\cite{Eroshenko1982, Eroshenko1988}. Storage of energy coincides with the intrusion step while its restoration coincide with the extrusion step (see figure ~\ref{cycle_eau}). Different behaviours  can be observed with either a low hysteresis, of interest for energy {storing} \cite{Eroshenko2001a}, or a high hysteresis of interest for damping applications \cite{Guillemot2012a, Suciu2003}. 



LHS can be seen as nanolaboratories to probe fluid transport, nucleation, phase changes, and more generally interfacial phenomena in nano or sub nanoconfinement. 
LHS allowed to demonstrated that the macroscopic Laplace-Washburn equation is still usable at nanoscale to predict intrusion pressure \cite{Grosu2014, Lefevre2004a}:

\begin{equation}
P_{int}=\frac{-2\gamma cos\theta}{R_p}
\end{equation}
whith $P_{int}$ the intrusion pressure, $\gamma$ the liquid-vapor surface tension, $\theta$ the contact angle between solid and liquid and $R_p$ the pore radius. 
It has also been brought to light that nucleation can be a bottleneck for the extrusion process when the pore radius is large or the temperature low  \cite{Guillemot2012b,Grosu2014}. 

Moving from pure water to weak electrolytes, Tzanis et al \cite{Tzanis2014} showed, using MFI-type zeolite as the media, that the presence of lithium chloride, magnesium chloride or sodium chloride increases both intrusion and extrusion pressures.
Qiao et al\cite{Kim2009, Liu2009} highlighted the same effect using different aqueous solutions and hydrophobic nanoporous materials with pressure increase which can reach $10 MPa$ for concentration as low as $0.5 mol.L^{-1}$ \cite{Han2008b}. This effect can be partly explained by surface tension changes due to the presence of ions in solutions \cite{Han2009, Kong2006}. However  we want to show in this communication that osmotic phenomena have also to be taken into account, precisely for very small pores within which ions can no longer enter.

We propose here to study a Metal Organic Frameworks (MOFs) called Zeolitic Imidazolate Framework 8 (ZIF-8) already presented combined with water elsewhere \cite{Ortiz2013}. The main advantage of such particles compared to silica-based one is the fact that they are naturally hydrophobic and, as a consequence, can be used in LHS without further chemical grafting which can lead to ageing issues or fluctuation in coverage rate between two batches. In this work we associated this nanoporous material with aqueous solutions of different ions. After a descriptive experimental section, we will present the results we obtained with ZIF-8/aqueous solutions systems. Then we will propose an explanation for these results on the basis of osmotic phenomena described, as far as possible, by a very simple perfect gas-type law.

\section{Experimental section}
\subsection{Nanoporous material and associated liquid}

\begin{figure}[htbp]
\begin{center}
\includegraphics[scale=0.3]{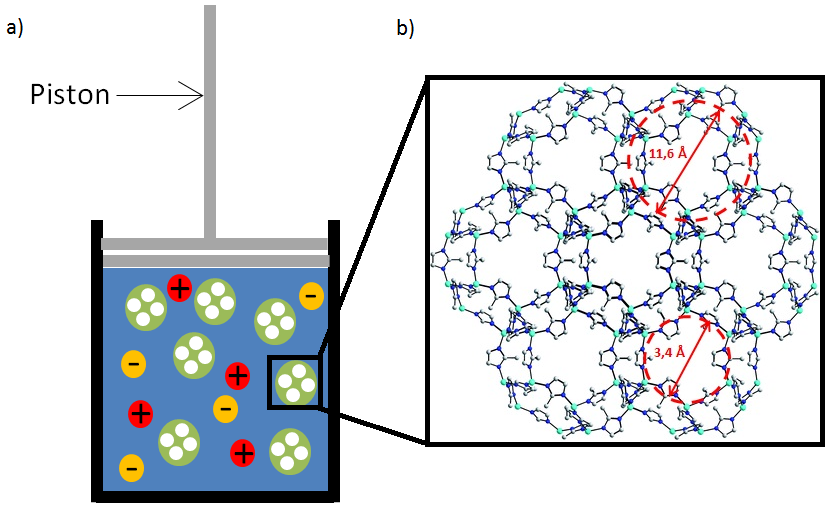} 
\end{center}
\caption{a) Principle of intrusion/extrusion tests. Cylinder contains ZIF-8 (in green) and electrolytes in solution. b) Schematic view of ZIF-8 structure.}
\label{ZIF}
\end{figure}

ZIF-8 (purchased from Aldrich) are MOFs with a sodalite morphology composed of cages with a diameter of 11.6 \AA~linked by small apertures of 3 \AA~\cite{Park2006}. This material has been characterized by Nitrogen adsorption experiment using an ASAP 2020 Physisorption Analyzer from Micromeritics. A specific areas $S_{BET}$ of $2000 m^2.g^{-1}$ has been determined by the Brunauer Emmett Teller \cite{Brunauer1938} method. A porous volume $V_p$ of $670 mm^3.g^{-1}$ has been calculated for a relative pressures of $0.9$. Using the $CuK\alpha$ radiation at 1.54 \AA~of a D8 advance Bruker diffractometer, we verified that the structure corresponds to a cubic space groupe $I\bar{4}3m$. A schematic view of the ZIF-8 structure is given on {figure~\ref{ZIF} b)}. For intrusion/extrusion tests, the material has been used without complementary treatment. 

Seven ions in solution were studied: lithium chloride $LiCl$, lithium iodide $LiI$, sodium chloride $NaCl$, sodium bromide $NaBr$, sodium iodide $NaI$, calcium chloride $CaCl_2$ and cesium chloride $CsCl$. All these solutions were prepared with powder salts (Sigma-Aldrich) and ultrapure water. In order to study the concentration effects, solutions with different concentrations were elaborated for $NaCl$ and $NaI$. 

\subsection{Intrusion/extrusion tests}

Intrusion/extrusion tests were achieved using a piston-cylinder device {(see figure~\ref{ZIF} a))}, $11.3$~mm in diameter, described by Guillemot et al \cite{Guillemot2012a} mounted on a traction/compression machine. Temperature and pressure inside the chamber as well as the piston displacement are measured during the experiment. We chose to study constant-speed cycles at $0.5 cm.min^{-1}$ with one second-step between two ramps. 
Experiments were led at temperatures ranging from $303$~K to $343$~K. Reference intrusion/extrusion cycles were also performed with ZIF-8/pure water. 
Cycles and associated characteristics are presented after substraction of the different compressibilities in order to focus on the intrusion/extrusion process.


\section{Results}

\begin{figure}[htbp]
\begin{center}
\includegraphics[width=\linewidth]{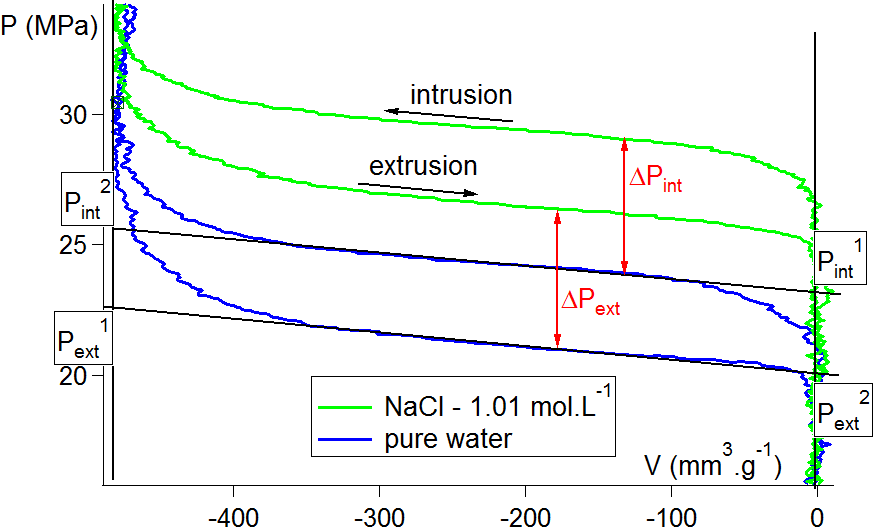} 
\end{center}
\caption{Hysteresis cycle of ZIF-8/water (in blue) and ZIF-8/NaCl - $1.01$ mol.L$^{-1}$ (in green) at $343 K$.}
\label{cycle_eau}
\end{figure}

Typical responses for ZIF-8/water and ZIF-8/$NaCl - 1.01 mol.L^{-1}$ couples are presented in figure~\ref{cycle_eau}. Intrusion (resp. extrusion) pressure is here defined as the mean of the intersection pressures between two vertical lines (corresponding to beginning and end of intrusion (resp.extrusion)) and a line tangential to intrusion (resp. extrusion) plateau. ZIF-8/water leads to a quite small hysteresis with flat plateaus for both intrusion and extrusion and an intruded volume of about $500 mm^3.g^{-1}$ slightly smaller than porous volume. 
Intrusion and extrusion pressures are respectively $25.2 \pm 0.2 MPa$ and $20.9 \pm 0.2 MPa$ for pure water system. The excess pressure between aqueous solution and pure water, $\Delta P_j= P^{aqueous~solution}_j-P^{pure~water}_j$, $j=int$ for intrusion and $j=ext$ for extrusion, are determined for each salt at $323$~K (see {Supplementary informations}).  

\begin{figure}[htbp]
\begin{center}
\includegraphics[width=\linewidth]{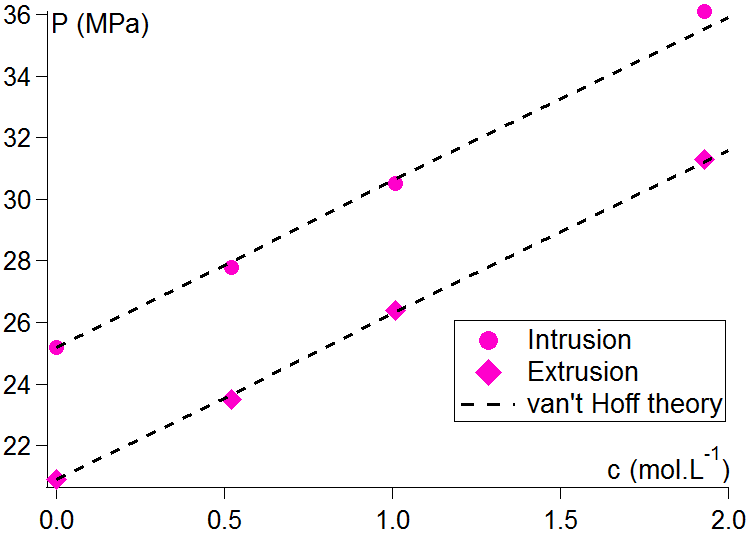} 
\end{center}
\caption{Intrusion and extrusion pressure as a function of concentration for NaCl solutions.}
\label{NaCl}
\end{figure}


{It can be seen on figure 2 that $NaCl - 1.01 mol.L^{-1}$ curve is obtained by a simple translation of the water curve. As a consequence, dissipated energy (which corresponds to the area between intrusion and extrusion plateaus) $E_d=2.2 \pm 0.4 J.g^{-1}$ is the same for both systems. Only stored energy (which corresponds to the area under the intrusion plateau) increases from $E_s^{water} =12.7 \pm 0.4$ J.g$^{-1}$ to $E_s^{NaCl - 1.01 mol.L^{-1}}=15.3 \pm 0.4$ J.g$^{-1}$.}

{As shown in figure~\ref{NaCl}, $P_{int}$ and $P_{ext}$ increase linearly with the concentration within the studied range of concentration for $NaCl$. And, as far as, intrusion and extrusion pressures increase by the same value for all these concentrations, $NaCl$ hysteresis are obtained by a simple translation to higher pressures of the pure water curve.}

{Surface tension of NaCl solutions is well-known \cite{Levin2001}. If we put aside contact angle effects, the expected surface tension for the different NaCl solutions under investigation should vary by less than 5\% leading to intrusion pressure variations of the same order. Nevertheless, the observed variations are higher than 10\% and goes to 50\% for the highest concentrated solution. Obviously, at such small scales, surface tension and macroscopic variations are probably not satisfying but such differences between theoritical and experimental values are surprising. Hereafter, we propose an interpretation of the experimental results on the basis of osmotic pressure and ion exclusion without using any nanoscopic argument or adjustable parameter.}

\section{Discussion}

As ions size and pores size are comparable, an ion exclusion process is plausible. If this is the case our experimental results should be explained by the means of osmotic pressure. 
In the case of LHS, the osmotic pressure corresponds to the additional pressure required to force the intrusion in the pores of an almost pure water solution extracted from the surrounding high concentration solution. 

\begin{figure}[htbp]
\begin{center}
\includegraphics[scale=0.3]{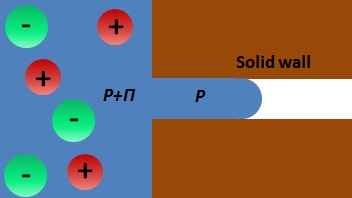} 
\end{center}
\caption{Schematic view of the ion exclusion during intrusion or extrusion leading to the appearance of osmotic pressure. The pressure inside (resp. outside) the pore is $P$ (resp. $P+\Pi$).}
\label{exclusion}
\end{figure}

In other words, it is the needed pressure to partially or totally filtrate the solution through pores. It has been demonstrated  that, in the case of a perfect diluted solution {where ions are totally excluded from one medium (such as shown on figure~\ref{exclusion})}, the osmotic pressure can be written following the van 't Hoff law\cite{VantHoff1995}:

\begin{equation}
\Pi=icRT
\end{equation}
where $\Pi$ is the osmotic pressure, $i$ is a coefficient corresponding to the number of ions in a salt (for instance $i=2$ for NaCl which will dissociate in one $Na^+$ and one $Cl^-$), $c$ is the molarity, $R$ is the perfect gas constant and $T$ is the absolute temperature.

\begin{figure}[htbp]
\begin{center}
\includegraphics[width=\linewidth]{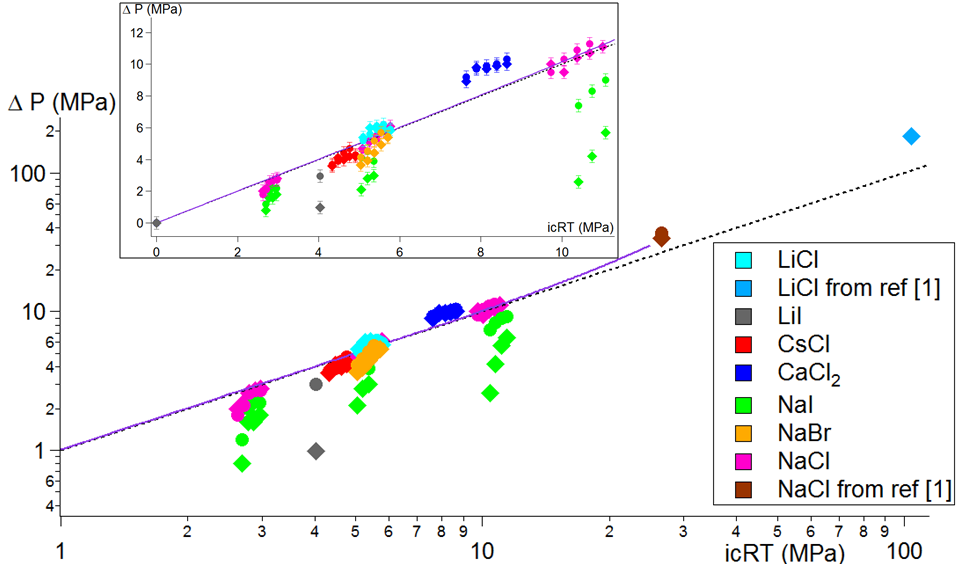} 
\end{center}
\caption{Validation of the osmotic pressure law. $\Delta P$ is represented as a function of theoretical $\Pi=icRT$ for intrusion ($o$) and extrusion ($\meddiamond$). van 't Hoff law (- - -) and a simulation from ref \cite{Luo2010} (\textcolor{Fuchsia}{$\relbar$}) are also reported. Top left frame is a magnification of the curve.}
\label{resume}
\end{figure}

Figure~\ref{resume} summarizes the results with experimental $\Delta P$ plotted according to $icRT$. As predicted from van 't Hoff law, points obtained from low concentration  $LiCl$, $NaCl$, $NaBr$ and $CsCl$ solutions align along a straight lined whose slope is 1.

At large concentration however van 't Hoff law is not correct anymore as interactions between ions are not negligible anymore compared to thermal fluctuations. The typical distance for which the electrostatic interactions are of the same order of magnitude is given by the Bjerrum length $\lambda_B$ :
\begin{equation}
\lambda_B \sim \frac{\mid q^+q^- \mid}{4 \pi \epsilon_0 \epsilon_r k_B T}
\end{equation}
where $q^+q^-$ is the product of positive and negative charges, $\epsilon_0$ the vacuum permitivity, $\epsilon_r$ the water dielectric constant and $k_B$ the Boltzmann constant. The Bjerrum length can be related to a number $n$ of ions per unit volume $n \sim \lambda_B^{-3}$ and the corresponding molarity:
\begin{equation}\label{clim}
c_{lim}=\frac{n}{\mathcal{N}_A}=\frac{1}{\mathcal{N}_A} \left( \frac{4 \pi \epsilon_0 \epsilon_r k_B T}{\mid q^+ q^- \mid} \right)^3
\end{equation}
with $\mathcal{N}_A$ the Avogadro number.

\begin{table}[htbp]
\begin{center}
\begin{tabular}{|c|c|c|c|}
\hline
Valence & $\lambda_B (nm)$ & $c_{lim} (mol.L^{-1})$ & $ic_{lim}RT (MPa)$\\ \hline
1 & 0.7 & 6.0 & 30 \\ \hline
2 & 1.4 & 0.1 & 0.5  \\ \hline
\end{tabular}
\end{center}
\caption{Bjerrum length and limit concentration for different valences at 303 K.}
 \label{Bjerrum}
\end{table}

Investigated concentrations for monovalent ions, at room temperature, are much lower than {the order of magnitude of this} limit molarity $c_{lim}=6.0$ mol.L$^{-1}$ (see table \ref{Bjerrum}). For concentrations close to or larger than $c_{lim}$ the osmotic pressure increases more rapidly than the linear van 't Hoff prediction\cite{VanGauwbergen1997, Prickett2008}. The full line curve in figure~\ref{resume} corresponds to molecular dynamic simulations carried out by Luo et al\cite{Luo2010}. It is worth to note, taken into account the trend of this curve, that osmotic pressure could be also a good candidate to explain pressures measured by Tzanis et al at the maximum $NaCl$ concentration with zeolites porous media (right hand side of figure \ref{resume})\cite{Tzanis2014}.

In equation \ref{clim} it can be seen that in addition to the concentration the valence of an ion with a three exponent has a major impact on $c_{lim}$. For a couple with one divalent ion and one monovalent ion such as $CaCl_2$, the limit molarity is $c_{lim}=0.1 mol.L^{-1}$. Studied concentrations are {much} larger than this concentration. As a result it is not surprising that experimental $\Delta P$ for $CaCl_2$ solution are larger than the van 't Hoff theoretical curve in figure \ref{resume}.
%

\begin{figure}[htbp]
\begin{center}
\includegraphics[width=\linewidth]{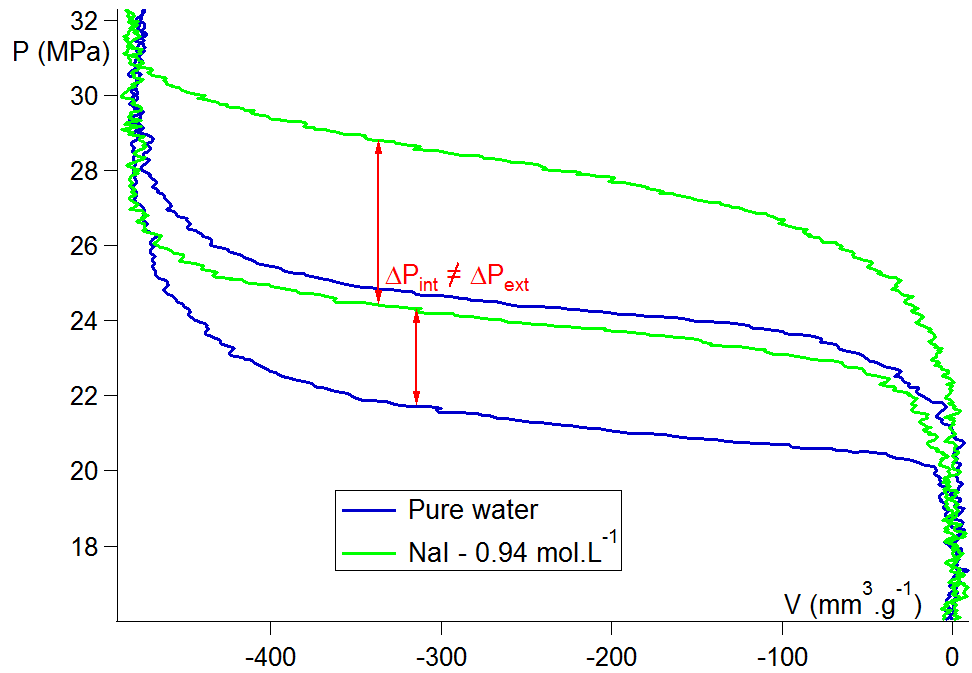} 
\end{center}
\caption{Hysteresis cycle of ZIF-8/water (in blue) and ZIF-8/NaI - $0.94 mol.L^{-1}$ (in green) at $343 K$.}
\label{cycle_NaI}
\end{figure}

Remains the particular case of $NaI$ and $LiI$ solutions. {Figure~\ref{cycle_NaI} represents the hysteresis cycle of $NaI - 0.94 mol.L^{-1}$ and pure water. This time, salt hysteresis is no more a simple translation of the pure water curve. In fact, intrusion pressure increases much more than extrusion one. As a consequence, both dissipated energy and stored energy are higher for iodide based system when compared with water.}

Excess pressures are slightly under the osmotic pressure curve showing only a partial filtration of ions. This results is not intuitive at all. In fact, ionic radius of iodide is much higher than chloride. Nevertheless this phenomenon can be explained by the fact {that iodide ions have a special affinity for hydrophobic surface~\cite{Huang2007}.}

Subsequently, if we put aside all effects linked to {$\gamma \cos\theta$} changes this notion and its variations with concentration becoming not relevent in such confined systems, it is possible to calculate the concentration inside the pores $c_{int}$ by applying the osmotic pressure law between a solution keeping the initial concentration (which is a rather good approximation as far as the intruded volume is very small regarding to the reservoir volume) and a solution of concentration $c_{int}$. Then, we can calculate the correponding volume occupied by one ion:

\begin{equation}
V_{ion}=\frac{1}{c_{int}\mathcal{N}_A}
\end{equation}

where $\mathcal{N}_A$ is the Avogadro constant. Finally, comparing to the volume of one ZIF-8 cage (about $1.56 nm^3$), we can conclude that about $1/4$ to $1/5$ of the cages are occupied by one ion. Table~\ref{occupation} sumarizes those results for $NaI$ solutions. Then, a very special behaviour is expected for nucleation because of this ion ordering. At that time, we are incapable to foresee the resulting macroscopic values and to compare it to experimental results for extrusion. Again, these $c_{int}$ values are to be considered with precaution, a macroscopic reasing being quite inaccurate at such small length scales.

\begin{table}[htbp]
\begin{center}
\begin{tabular}{|l|c|c|}
\hline
Aqueous solution & $c_{int} (mol.L^{-1})$ & $n_{occ}$ \\ \hline
$NaI - 0~mol.L^{-1}$ & 0 & 0  \\ \hline
$NaI - 0.52~mol.L^{-1}$ & $0.15 \pm 0.08$ & $0.07 \pm 0.04$  \\ \hline
$NaI - 0.94~mol.L^{-1}$ & $0.18 \pm 0.08$ & $0.09 \pm 0.04$  \\ \hline
$NaI - 2.0~mol.L^{-1}$ & $0.44 \pm 0.08$ & $0.21 \pm 0.04$  \\ \hline
${LiI - 0.75~mol.L^{-1}}$ & ${0.14 \pm 0.08}$ & ${0.07 \pm 0.04}$ \\ \hline
\end{tabular}
\end{center}
\caption{Intruded concentration, corresponding volume occupied by one ion and ion number by ZIF-8 cage $n_{occ}$ for the different $NaI$ solutions. }
 \label{occupation}
\end{table}

\section{Conclusions}

Here was described the influence of different ions over intrusion and extrusion pressures of a specific LHS using ZIF-8 as the porous material. The increase of intrusion and extrusion pressures just as stored energy with increasing ion concentration has been highlighted. It has been demonstrated that for diluted solution, a very simple osmotic pressure law can explain the observed phenomena by analogy with perfect gas. This law permits to explain both concentration and temperature behaviours observed on the different LHS. Moreover, contrary to intuition, we showed that it is easier for { iodide ions to enter the channels than for smaller ones as far as they show a very specific affinity for hydrophobic surfaces}. It has also been evidenced that this very simple law is only valid till a limit concentration which is very dependent on ion valence. Above this concentration, the intrusion and extrusion pressures seem to increase more quickly with concentration than what is predicted by van 't Hoff law.

\begin{acknowledgement}

The authors kindly thank Airbus Defence and Space and the French Direction Generale de l'Armement which support us in our researches.

\end{acknowledgement}




\bibliography{Publi}

\end{document}